\documentclass[pra,aps,twocolumn,showpacs,superscriptaddress,amsmath,amssymb]{revtex4-1}
\usepackage{color}

\usepackage{amsmath,graphicx}

\begin{document}
\def\tr{\rm{Tr}}
\def\la{{\langle}}
\def\ra{{\rangle}}
\def\a{{\alpha}}
\def\e{\epsilon}
\def\q{\quad}
\def\w{\tilde{W}}
\def\t{\tilde{t}}
\def\a{\hat{A}}
\def\h{\hat{H}}
\def\E{\mathcal{E}}
\def\p{\hat{P}}
\def\u{\hat{U}}
\def\n{\hat{n}}
\def\j{\hat{j}}
\def\s{\hat{S}}
\def\le{\leftarrow}

\title{Quantum law of rare events for systems with Bose-Einstein statistics
}
%
%
\author {D.  Sokolovski}
\affiliation{Departmento de Qu\'imica-F\'isica, Universidad del Pa\' is Vasco, UPV/EHU, Leioa, Spain}
\affiliation{IKERBASQUE, Basque Foundation for Science, E-48011 Bilbao, Spain}

\date{\today}
\begin{abstract}
In classical physics the  joint probability of a number of individually rare independent events is  given by the Poisson distribution. It describes, for example,
 unidirectional
  transfer of population between the densely and sparsely populated states of a classical two-state system. We derive a quantum version of the law for a large number of non-interacting systems (particles) obeying Bose-Einstein statistics. The classical low is significantly modified by quantum interference, which allows, among other effects, for the counter flow of particles back into the densely populated state. Suggested observation of this classically forbidden counter flow effect can be achieved with modern laser-based techniques used for manipulating and trapping of cold atoms.





\end{abstract}

%
%
\maketitle
%
%
%
%
%
%
%
In classical physics and statistics, probability for a number of individually rare events is universally given by the Poisson distribution (see, for instance, \cite{Pois}).
For example, it is obeyed by a classical gas escaping into an empty space through a penetrable membrane. 
With the number of atoms $N$ large, and the transition probability made proportionally small, the number of escaped atoms is governed by the Poisson law,  with the number of atoms recaptured into the original reservoir vanishing as $N\to \infty$. The validity of the Poisson distribution depends on that one can, in principle, know not only how many but also which of the atoms have escaped. 
Quantum mechanics offers a different possibility: for identical particles one is allowed to know only the number of the escapees, and not their identities.
While it is well known that both Fermi-Dirac and Bose-Einstein symmetries of a wave function may lead to non-poissonian effects in the full counting statistics of otherwise independent particles \cite{NPT-2}-\cite{NPT4}, the failure of the Poisson law in the limit of rare events is less obvious. 
The subject of this Letter is the general question of what replaces the classical Poisson law in a quantum situation where only the total number of rare events, but not their individual details, can be observed?.
\newline
 We specify to the case of many non-interacting bosons, each of which may occupy one of the two available states.
Such systems are also of practical interest, e.g., for their potential applications as detectors.
 For example, if the transmission amplitude between two connected cavities is influenced by a passing particle, the change observed  in the  photonic  current would announce the particle's arrival. In a similar way, atomic current of a weakly interacting  Bose-Einstein condensate (BEC) trapped in a double- or multi-well potential (see Fig.1)  can be used to gain information about the state of a qubit coupled to the BEC \cite{MIRR}-\cite{DS2}. Detailed analysis of the work of such hybrid bosonic devices must take into account in which manner, and how frequently, the bosonic sub-system  is observed, and will be given elsewhere.  
\newline 
We note that the problem is fundamentally different from that of the frequently studied coined quantum walk \cite{WALK}, where interference between virtual paths available to a single particle modifies the classical Gaussian distribution. In our case, modification of the classical law is a many-body effect, specific to the Bose-Einstein statistics.

\begin{figure}
\includegraphics[width=.9\linewidth]{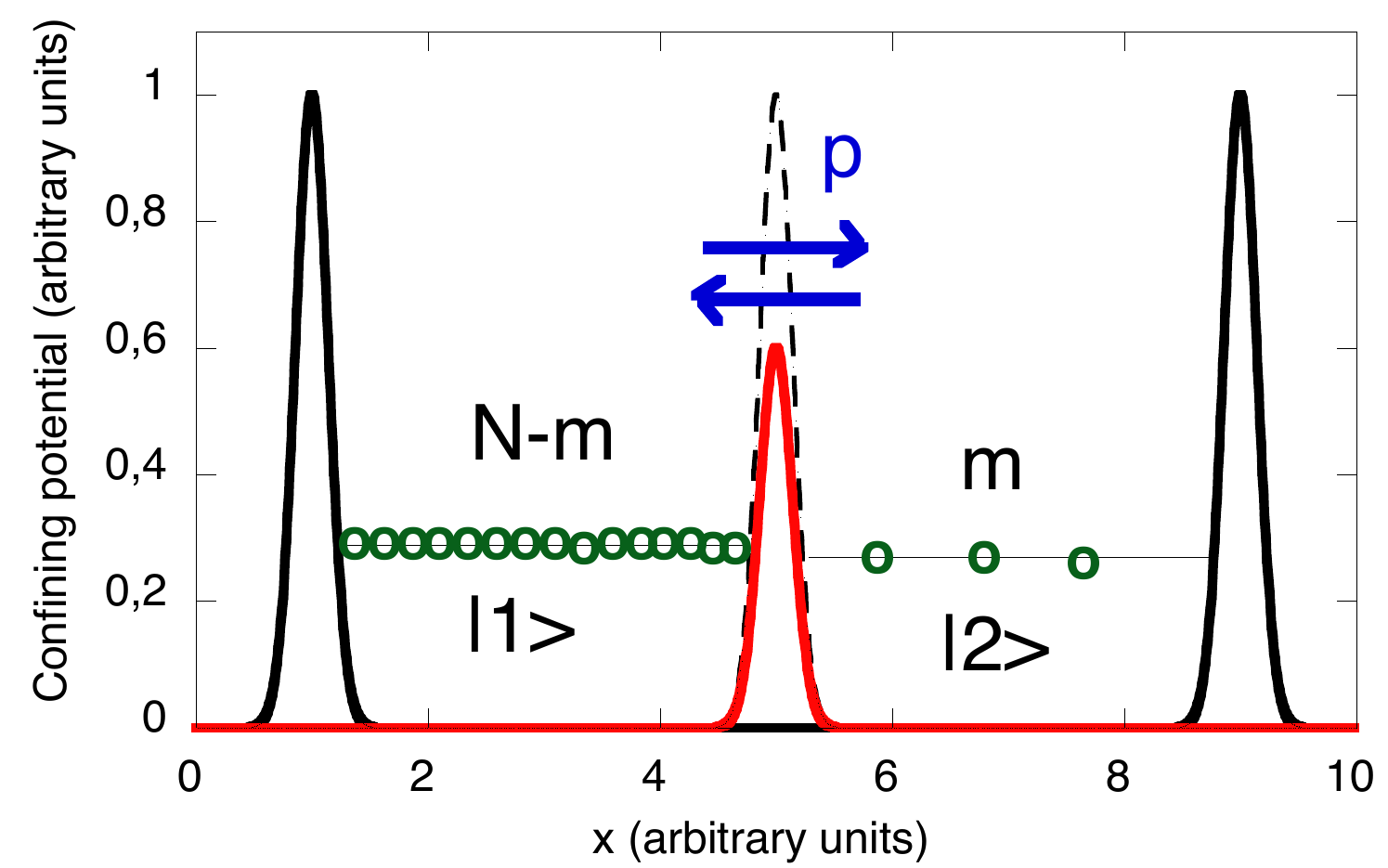}
  \caption{(Color online)  Double-well trap containing $N$ atoms. The central barrier is lowered to allow tunnelling between the states $|1\ra$ and $|2\ra$.}
\end{figure} 

We start by constructing transition amplitudes for a single particle, which can occupy one of the two levels in an asymmetric double well potential. In terms of the Pauli's matrices, the Hamiltonian reads
\begin{eqnarray}\label{1h}
\h=\epsilon \sigma_z + \xi \sigma_x  + \eta\sigma_y,
\end{eqnarray}
where the spin states $|1\ra$ and $|2\ra$, aligned up and down the $z$-axis, represent an atom in the right and left state, respectively, $\epsilon$ is the difference between the energies of the states, and $\xi$ and $\eta$ together define the tunnelling matrix element, $T=\xi+i\eta$.
For the evolution operator $\u(t)=\exp(-i\h t)$ we have
\begin{eqnarray}\label{2h}
\u(t)= Icos(\omega t)-i\omega^{-1}[\epsilon \sigma_z + \xi\sigma_x  + \eta \sigma_y]\sin(\omega t) ,   
\end{eqnarray} 
with $\omega= \sqrt{\epsilon^2+\xi^2+\eta^2}$, 
and its matrix elements are conveniently written as (a star indicates complex conjugate)
\begin{eqnarray}\label{6h}
{U}_{11}=\sqrt{1-p}\exp(i\alpha)={U}_{22}^*
\\ \nonumber
{U}_{12}=\sqrt{p}\exp(i\beta)=-
{U}_{21}^*,
\end{eqnarray} 
where, with our choice of the basis, $p(t)=(\xi^2+\eta^2)\sin^2(\omega t) /\omega^2$
is one-particle transition probability, $\alpha(t)=
-\arctan[\epsilon\tan(\omega t)/\omega]$ and $\beta=-\pi/2$.

For a total of $N$ particles, we wish to evaluate the transition probabilities $p^{N}_{m'\le m}(t)$ for starting with $m$ particles in the state $|1\ra$ and ending, after a time $t$, with $m'$ particles in the same state.
It is instructive to begin with a brief discussion of the case where all particles are considered distinguishable. 
The problem is equivalent to a classical $N$-coin one: given that each coin changes its state with a probability  $p$, and $m$ coins initially oriented heads up, what is the probability to have $m'$ heads up after each coin has been tossed once? The result can be achieved by moving $\nu$ coins from tails to heads, and $\mu$ coins from heads to tails, provided $\nu-\mu=m'-m$. Summing the corresponding probabilities, while taking into account the number of ways to choose the coins which change their state, yields
\begin{eqnarray}\label{1a}
p^N_{m'\le m}=\sum_{\mu=0}^m\sum_{\nu=0}^{N-m}C^m_\mu C^{N-m}_\nu \\ \nonumber
\times p^{\nu+\mu}(1-p)^{N-\mu-\nu}\delta_{\nu-\mu, m'-m},
\end{eqnarray}
where $C^k_l\equiv \frac{k!}{l!(k-l)!}$ is the binomial coefficient, and $\delta_{mn}$ is the Kronecker delta. 
\begin{figure}
\includegraphics[width=0.8\linewidth]{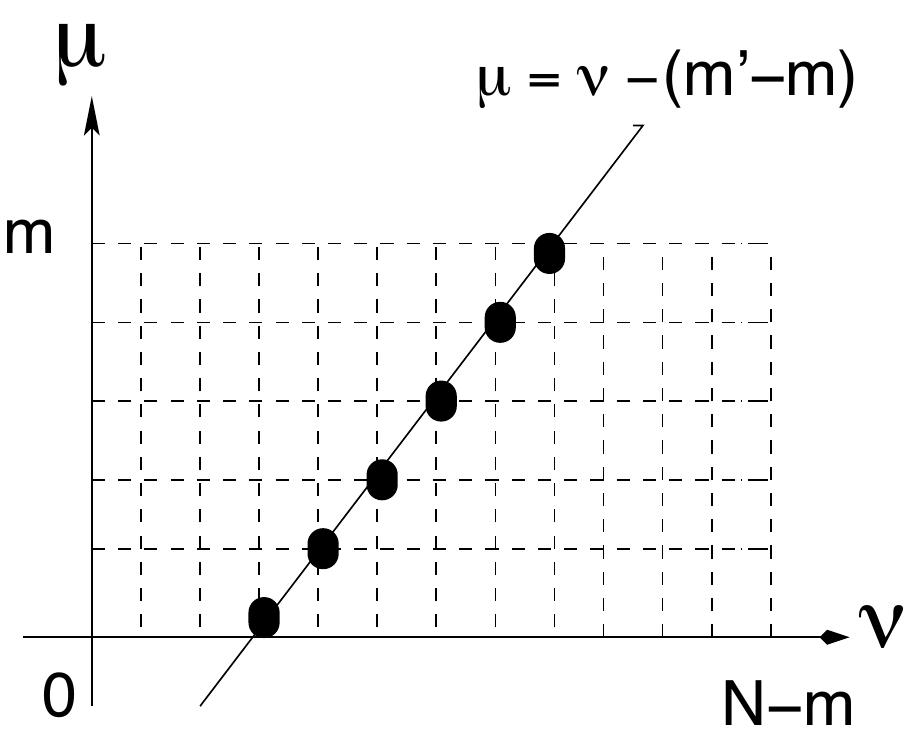}
  \caption{ A diagram showing the region of summation in Eqs. (\ref{1a}) and (\ref{4g}) (filled dots). Each dot contributes $C^m_\mu C^{N-m}_\nu p^{\nu+\mu}(1-p)^{N-\mu-\nu}$  for distinguishable particles, and $C^m_\mu C^{N-m}_\nu(-1)^{\mu} U_{12}^{\nu+\mu}U_{11}^{m-\mu}U_{22}^{N-\mu-\mu}$ for identical bosons.}
\end{figure} 
Depending on $N$,$m$ and $m'$, the sum in Eq.(\ref{1a}) may contain a different number of terms, corresponding to the number of 'pathways' connecting the initial and final states (filled dots in Fig.Á2). In the rare events (RE) limit
\begin{eqnarray}\label{2a}
N\to \infty, \q p\to w/N,
\end{eqnarray}
it is sufficient to retain only the leading $\mu=0$ terms (the lowest dot in the diagram in Fig.2) in (\ref{1a}). 
Using the relation $lim_{N\to\infty} C^N_m=N^m/m!$ for the remaining binomial coefficient, yields the expected Poisson distribution,
\begin{eqnarray}\label{1ir}
\lim_{N\to \infty}p^N_{m'\le m}= \begin{cases}w^q \exp(-w)/q! \q q\equiv m'-m\ge 0\\
0\q\q\q\q\q\q\q q<0.\end{cases}
\end{eqnarray}
with intuitively appealing properties. 
Indeed, in the case of a symmetric trap, $\epsilon=0$, reducing the transition probability in (\ref{2a}) also makes the Rabi period $2\pi/\omega$ after which the system must return to its initial state, extremely large. Now, for all $t<< 2\pi/\omega$, the evolution can be considered approximately  irreversible, with number of particles $q$ escaping into the right trap independent of number of particles, $m$,  already there. 
Low probability of each individual event, and much lower population in the right well make re-crossings from right to left extemely unlikely (see Fig.3).
 In particular, after detecting $m$ particles in the right well, one never finds 
it  empty again,  as the probability $p^N_{0\le m}$ (the left upper corner in the diagram in Fig. 2), vanishes as $(w/N)^m\exp(-w)$. One might expect a similar argument to be also valid should distinct particles be replaced with non-interacting bosons. Next we will show that this is not the case.
\begin{figure}
\includegraphics[width=1.\linewidth]{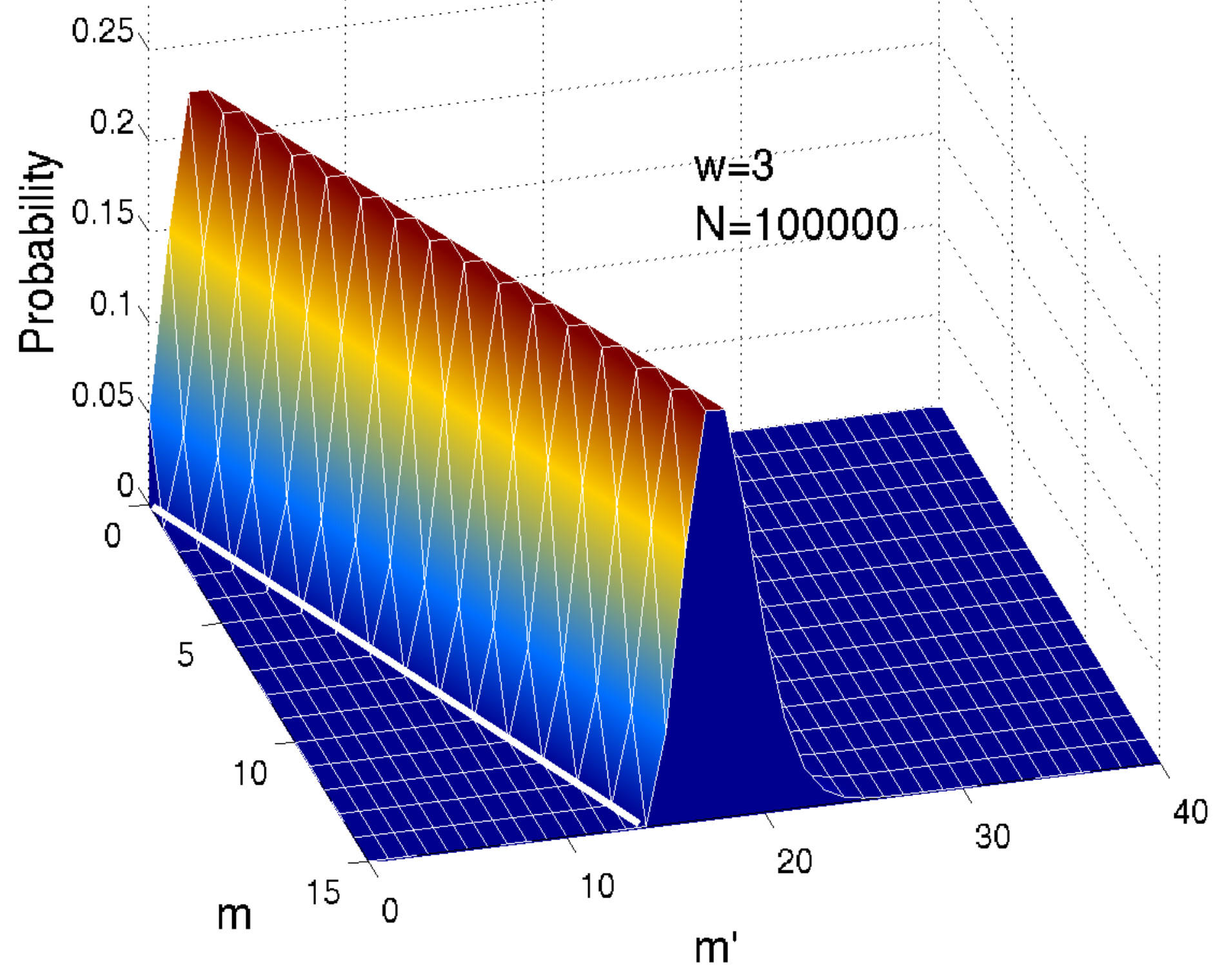}
  \caption{(Color online)  Poissonian probability $p^N_{m'\le m}$ for distinguishable particles, as given by Eq.(\ref{1a}),  for $N=10^5$ and $w=3$. The white line marks $m=m'$.}
\end{figure} 

For identical bosons we have a quantum version of the $N$-coin problem: after a toss each coin changes its state from $|i\ra$ to $|j\ra$, $i,j=1,2$ with the probability amplitude $U_{ji}$, and we must sum amplitudes rather than probabilities over all pathways leading to the same final state.
The state of the system with any $m$ coins displaying heads is given by a symmetrised wave function 
\begin{eqnarray}\label{1}
|m,N\ra = (C^N_m)^{-1/2}\sum \prod_{j=1}^N |i_j\ra_j,\q i_j=1,2,
\end{eqnarray}
where $|i\ra_j$, $i=1,2$ denotes the state of the $j$-th particle, and the sum is over $C^N_m$  different ways to ascribe to $m$  of the $N$ indices $i_j$
the value of $2$, and to the remaining $N-m$ ones the value of $1$.
After all coins are tossed once each individual term in the sum
 of Eq.(\ref{1})
contributes to the amplitude to have $m'$ heads up a quantity
\begin{eqnarray}\label{2}\nonumber
f(m'\leftarrow m,N) = (C^N_m)^{-1/2}(C^N_{m'})^{-1/2}\q\q\q\q\\
\times \sum_{\mu=0}^{m}\sum_{\nu=0}^{N-m}
C^m_\mu C^{N-m}_\nu{U}_{12}^{\nu}{U}_{21}^{\mu}{U}_{11}^{m-\mu}
{U}_{22}^{N-m-\nu}\q\q\q
 \\ \nonumber
\times \delta_{m'-m,\nu-\mu},\q\q\q\q
\end{eqnarray}
with the region of summation illustrated in Fig.2.
Since Eq.(\ref{1}) contains $C^N_m$ such terms, the probability  to have 
$m'$ heads up after the toss is
$P^N_{m'm} = (C^N_m)^2|f(m'\leftarrow m,N)|^2$,  $m, m' =0,1,...N$,
which, with the help of Eq.(\ref{6h}), can be expressed in terms of the Jacobi polynomials
$\mathcal{P}_n^{(\alpha,\beta)}(x)$  \cite{AST}-\cite{JAC} 
\begin{eqnarray}\label{4g}
P^N_{m'\le m} = 
\frac{m!(N-m)!}{m'!(N-m')!}
p^{m'-m}(1-p)^{N-m'+m}\q\q\q\q\q\\ \nonumber
\times|\mathcal{P}_m^{(N-m'-m,m'-m)}(2p-1) |^2.\q\q\q
\end{eqnarray}
As in the case of distinguishable particles, $P^N_{m'\le m}$ depends only on the one-particle transition probability  $p$, and not of the phases $\alpha$ and $\beta$ of the matrix elements of
of $ U_{ij}$ in Eq.(\ref{6h}) \cite{FOOT1}. We note also that in the special case of tunnelling into an initially empty well, $m=0$, there is only one pathway (moving exactly $m$ particles from left to right), and
transition probabilities for distinguishable particles and identical bosons coincide, 
\begin{eqnarray}\label{6}
P^N_{m'\le 0} = p^N_{m'\le 0}= C^{N}_{m'}p^{m'}(1-p)^{N-m'},\q\q\q\q\q\q
\end{eqnarray}
as was pointed out earlier in the Refs. \cite{DS1} and \cite{DS2}. 
\newline
More interesting, however, are the transitions affected by the interference effects which, as we will demonstrate, persist even in the RE limit (\ref{2a}). Indeed, since the sum in Eq.(\ref{2}) contains $\sqrt{p}$ rather than $p$, the restriction to only $\mu=0$ terms is no longer justified. Thus, after taking the limit (\ref{2a}),  we have ($q=m'-m$)
\begin{eqnarray}\label{2ir}
\lim_{N\to \infty}P^N_{m'\le m}
=\
w^{q}\exp(-w)\q\q\\ \nonumber
 \times{|}\sum_{\mu=max[0,-q]}^{m}
 \frac{\sqrt{m'!m!}(-w)^{\mu}}{\mu!(m-\mu)!(q+\mu)!}|^2.
\end{eqnarray}
Equation (\ref{2ir}), which is our central result \cite{FOOT}, replaces the classical Poisson law (\ref{1ir}) for non-interacting identical bosons. Some of its properties are counterintuitive, as is shown in Fig.4. Firstly,
unlike the Poisson distribution (\ref{1ir}),  $P^N_{m'm}$ of Eq.(\ref{2ir}) is highly structured, as a result of the interference between the pathways. 
Secondly, it allows for total or partial recapture of the few particles initially held in the right well back into the densely populated left well, contrary to the simple argument based on improbability of such an event.
\begin{figure}
\includegraphics[width=1.\linewidth]{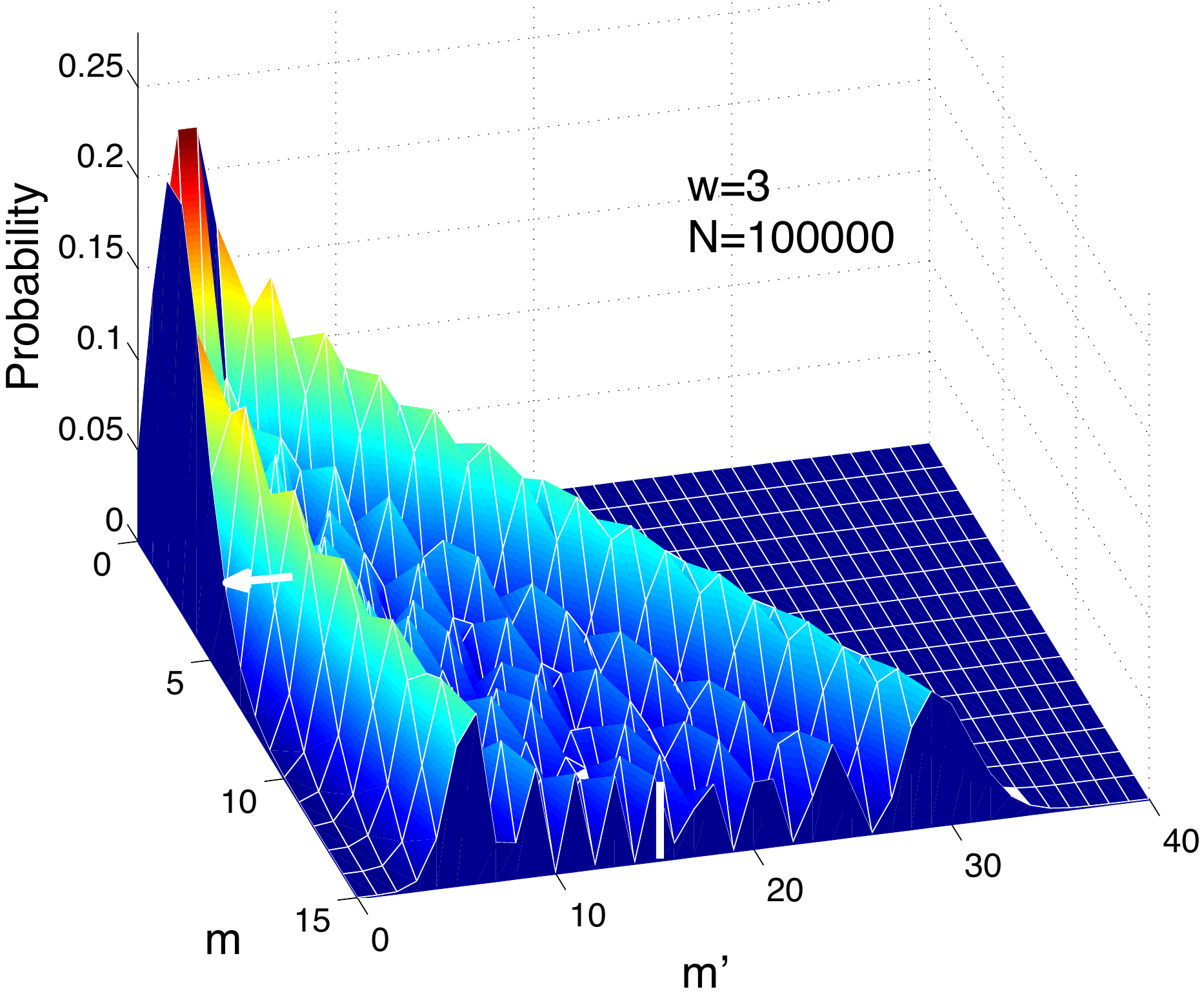}
  \caption{(Color online)  Non-poissonian probability $P^N_{m'\le m}$ for identical bosons, as given by Eq.(\ref{4g}),  for $N=10^5$ and $w=3$. The section of the surface indicate by the arrow corresponds to $P^N_{0\le m}$ also shown in Fig.4.}
\end{figure} 
 
 The probability for all $m$ bosons to cross into the left well, $P^N_{0\le m}$, contains only one term in the sum (\ref{2ir}) [$(\mu=m,\nu=0)$ in the diagram in Fig. 1], 
 \begin{eqnarray}\label{1c}
P^N_{0\le m}=w^m\exp(-w)/m!.
\end{eqnarray}
\begin{figure}
\includegraphics[width=1.\linewidth]{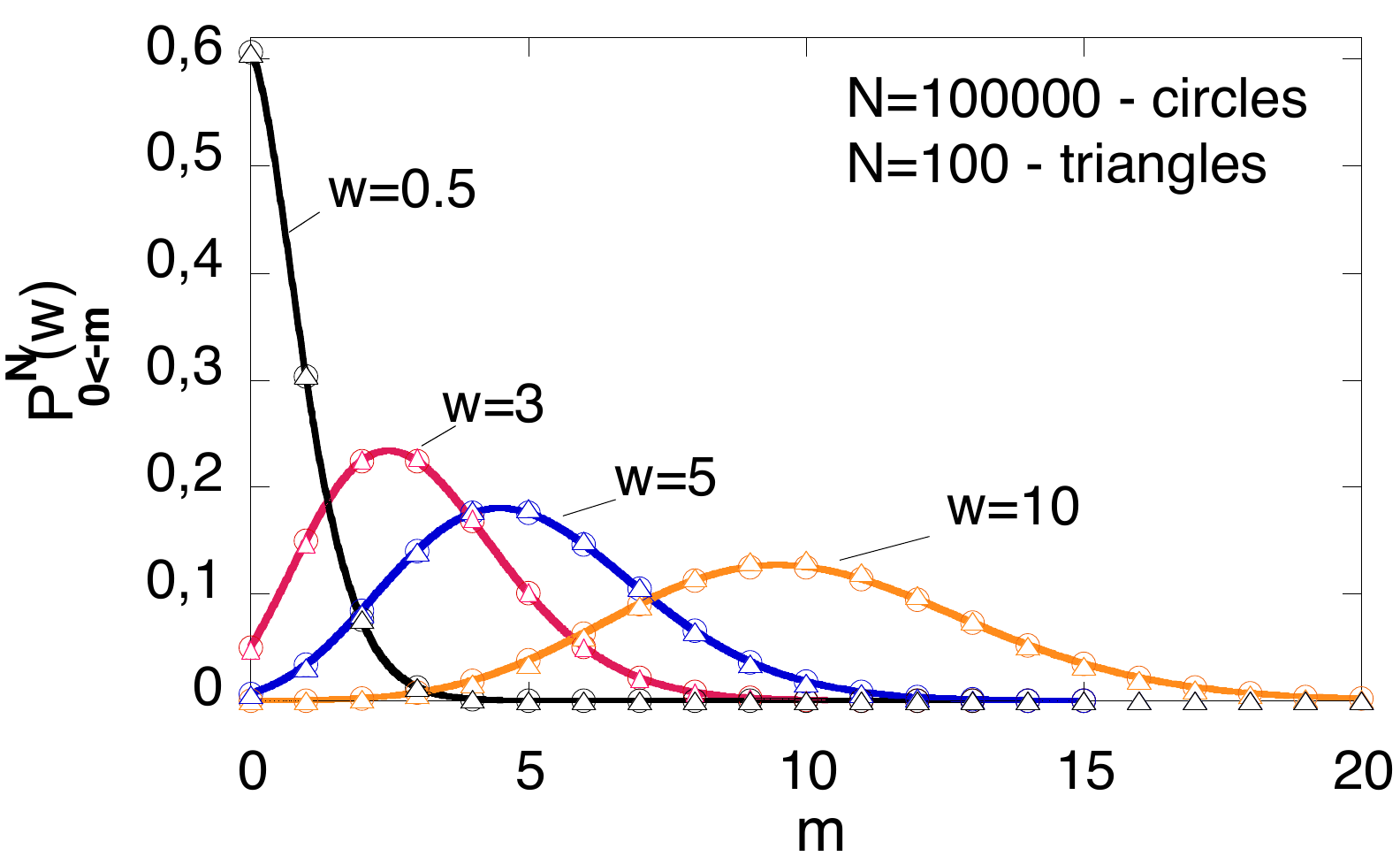}
\label{fcs}
 \caption{(Color online)  The probability $P^N_{0\le m}$ in Eq.(\ref{4g}) that all $m$ particles initially in the right well, $m<<N$, $N>>1$, would cross to the left, leaving the right well empty.
 The solid lines are the corresponding Poisson distributions (\ref{1c})}
\end{figure} 
As a function of the number of recaptured atoms $m$, it is a Poisson distribution shown in Fig.5 (apparently so ubiquitous that after having been evicted from one part of this paper, it immediately reappears in another, albeit in a different context). The recapture process exhibits certain resonance-like behaviour. The number of particles most likely to be readmitted to the left well, $m\approx w$, equals the mean number of 
distinguishable particles crossing into the right well under the same conditions. 
For $m>0$, there are two interfering scenarios leading to only one particle being left in the right well [points $(\mu=m -1,\nu=0)$ and  $(\mu=m,\nu=1)$ in Fig.2], and the corresponding probability is bimodal as shown in Fig. 4. Similarly, the probability $P^N_{m\le m}$
to retain the same number of atoms in the right well builds up from $m+1$ interfering terms and 
also shows an oscillatory pattern (see Fig.6).

To conclude, we suggest a simple experimental setup to test the re-capture property of the bosonic distribution. Using the available laser technology \cite{NPT-1} one can create quasi-one-dimensional box-trap with two strong endcap lasers providing the potential walls shown in Fig.1.
The box is divided in two by adding a third laser in the middle, and the left well is populated with a large number $N$ of weakly interacting atoms, while, say, three atoms are introduced into the right well. Following this, the middle laser beam is slightly weakened to allow transfer of atoms between the wells. It is restored after a time $\tau$, such that $p(\tau) \approx \xi^2\tau^2\approx 
3/N$, and the number of atoms in the right well is measured, e.g., by a technique described in \cite {DETECT}.
Then, no matter how large is $N$, 
the well is found empty with the probability $P^N_{0\le m}\approx  0.23$  (c.f., Fig. 5), i.e., in just under a quarter of all cases.
\begin{figure}
\includegraphics[width=1.\linewidth]{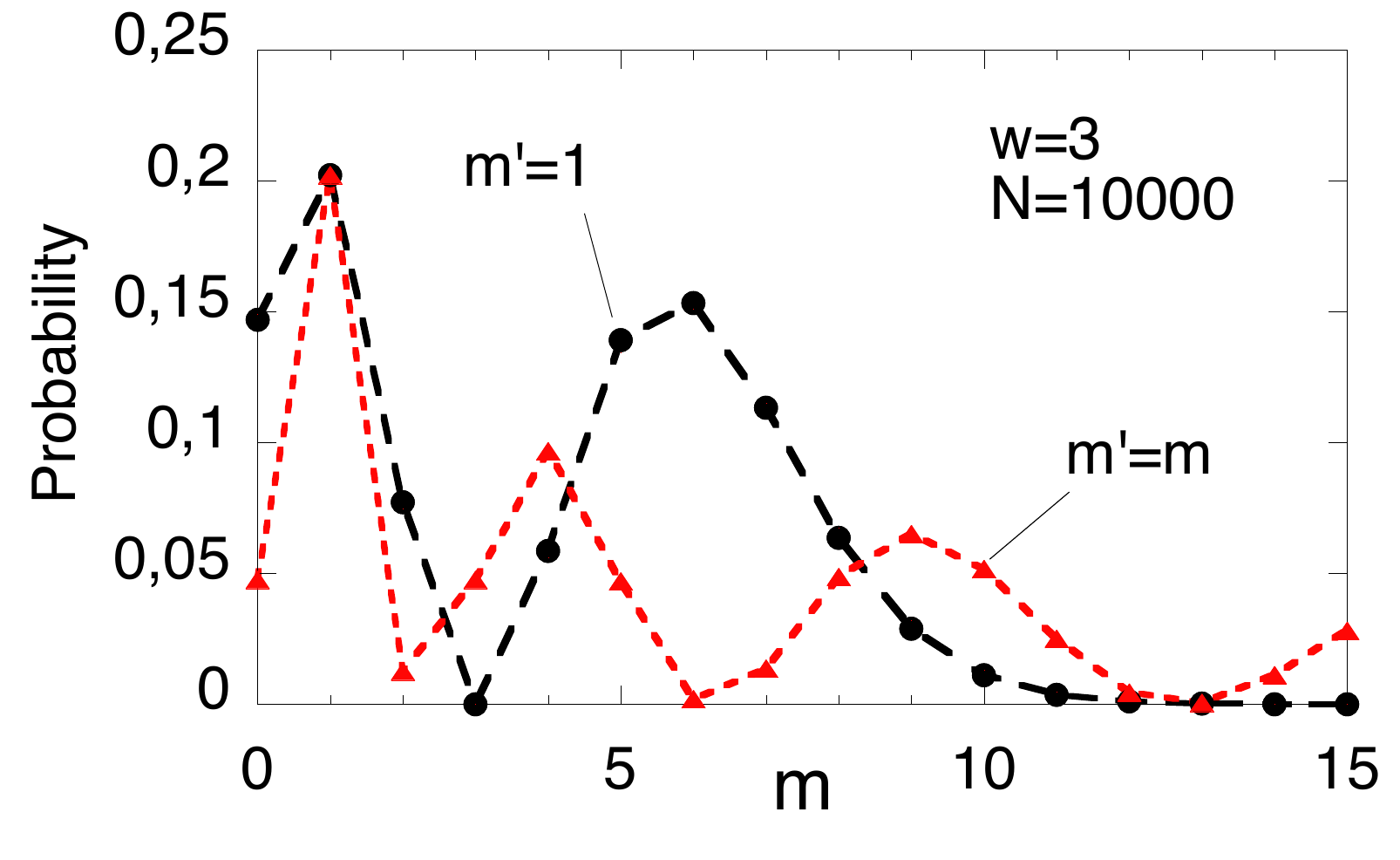}
\label{fcs}
 \caption{(Color online)  The probabilities $P^N_{1\le m}$ to end up with just one atom in the right well (circles, long dashed), and $P^N_{m\le m}$, to leave the population of the right well unchanged (triangles, dashed), as functions of $m$,}
\end{figure} 

In summary, it is shown that for non- or weakly interacting bosons quantum interference between different scenarios leading to the same final state modifies the classical Poisson law of rare events, and leads to significant observable effects not present in classical statistics.
\acknowledgements

We acknowledge support of the Basque Government (Grant No. IT-472-10), and the Ministry of Science and Innovation of Spain (Grant No. FIS2009-12773-C02-01). 


\begin{thebibliography}{999}
\bibitem{FOOT1} The interference effects discussed below can be seen as coming from  the minus sign in the second of the  Eqs.(\ref{6h}).
\bibitem{FOOT} Equation (\ref{2ir}) can be simplified further by exploring the asymptotic behaviour of Jacobi polynomials in the limit (\ref{2a}) e.g., by the methods of Ref. \cite{JAC}, but we will not 
pursue it further.
\end{thebibliography}
\end{document}